\input{aipcheck}

\documentclass[
    ,final            
  ]
  {aipproc}

\usepackage{amssymb}
\layoutstyle{8x11single}

\def\nn{\nonumber\\}

\def\lb{\left(}
\def\rb{\right)}
\def\hmu{\hat\mu}
\def\L{\ln\frac{\hat\Lambda}{2}}
\def\Lg{\ln\frac{\hat\Lambda_g}{2}}

\def\be{\begin{eqnarray}}
\def\ee{\end{eqnarray}}
\def\del{\partial}
\def\[{\left[}
\def\]{\right]}  
\def\lb{\left(}
\def\rb{\right)}
\def\hmu{\hat\mu}
\def\nn{\nonumber\\}

\def\Lg{\ln\frac{\hat\Lambda_g}{2}}
\newcommand{\beq}{\begin{equation}}
\newcommand{\eeq}{\end{equation}}
\newcommand{\bqa}{\begin{eqnarray}}
\newcommand{\eqa}{\end{eqnarray}}

\begin{document}

\title{Equation of State for QCD at finite temperature and density. Resummation 
versus lattice data}

\classification{64.30.Ef, 12.38.Mh, 12.38.Gc.}
\keywords      {QCD equation of state, Quark-gluon plasma, 
Hard-thermal-loop perturbation theory, Lattice QCD calculations.}

\author{Jens O. Andersen\hspace{-1mm}~\footnote{Speaker.}\hspace{1mm}}{
  address={Department of Physics, 
Norwegian University of Science and Technology, N-7491 Trondheim, 
Norway}
}
\author{Najmul Haque}{
  address={Theory Division, Saha Institute of Nuclear Physics,
1/AF Bidhannagar, Kolkata-700064, India}
}

\author{Munshi G. Mustafa}{
  address={Theory Division, Saha Institute of Nuclear Physics,
1/AF Bidhannagar, Kolkata-700064, India}
}
\author{Michael Strickland}{
  address={Department of Physics, Kent State University, Kent, Ohio 44242, 
United States}
}

\author{Nan Su}{
  address={Fakult\"at f\"ur Physik, Universit\"at Bielefeld, 33615 Bielefeld, 
Germany}
}


\begin{abstract}
The perturbative series for finite-temperature field theories has very poor 
convergence properties and one needs a way to reorganize it. In this talk, 
I review two ways of reorganizing the perturbative series for field theories at 
finite temperature and chemical potential, namely hard-thermal-loop perturbation
theory (HTLpt) and dimensional reduction (DR). I will present results for the 
pressure, trace anomaly, speed of sound and 
the quark susceptibilities from a 3-loop HTLpt calculation and for 
the quark susceptibilities using DR at four loops. A careful comparison with 
available lattice data shows good agreement for a number of physical quantities.
\end{abstract}

\maketitle


\section{Introduction}
In this talk I would like to discuss a problem that has been around for
a couple of decades namely the poor convergence of the perturbative
series of the thermodynamic functions of hot and dense QCD.
The weak-coupling expansion of the QCD pressure has a very long story
going back to the late 1970s~\cite{shuryak,chin,kapusta79,toimela,arnoldzhai1,arnoldzhai2,zhaikastening,eric3} and the interest has in part been spurred
by application to quark-gluon plasma phenomenology 
in heavy-ion collisions. 
The calculational frontier has been pushed
to order $g^6\ln(g)$ at zero chemical potential~\cite{kajantie}
and finite chemical potential~\cite{aleksi0,aleksi,ibb}.

\begin{figure}[htb]
  \includegraphics[height=.34\textheight]{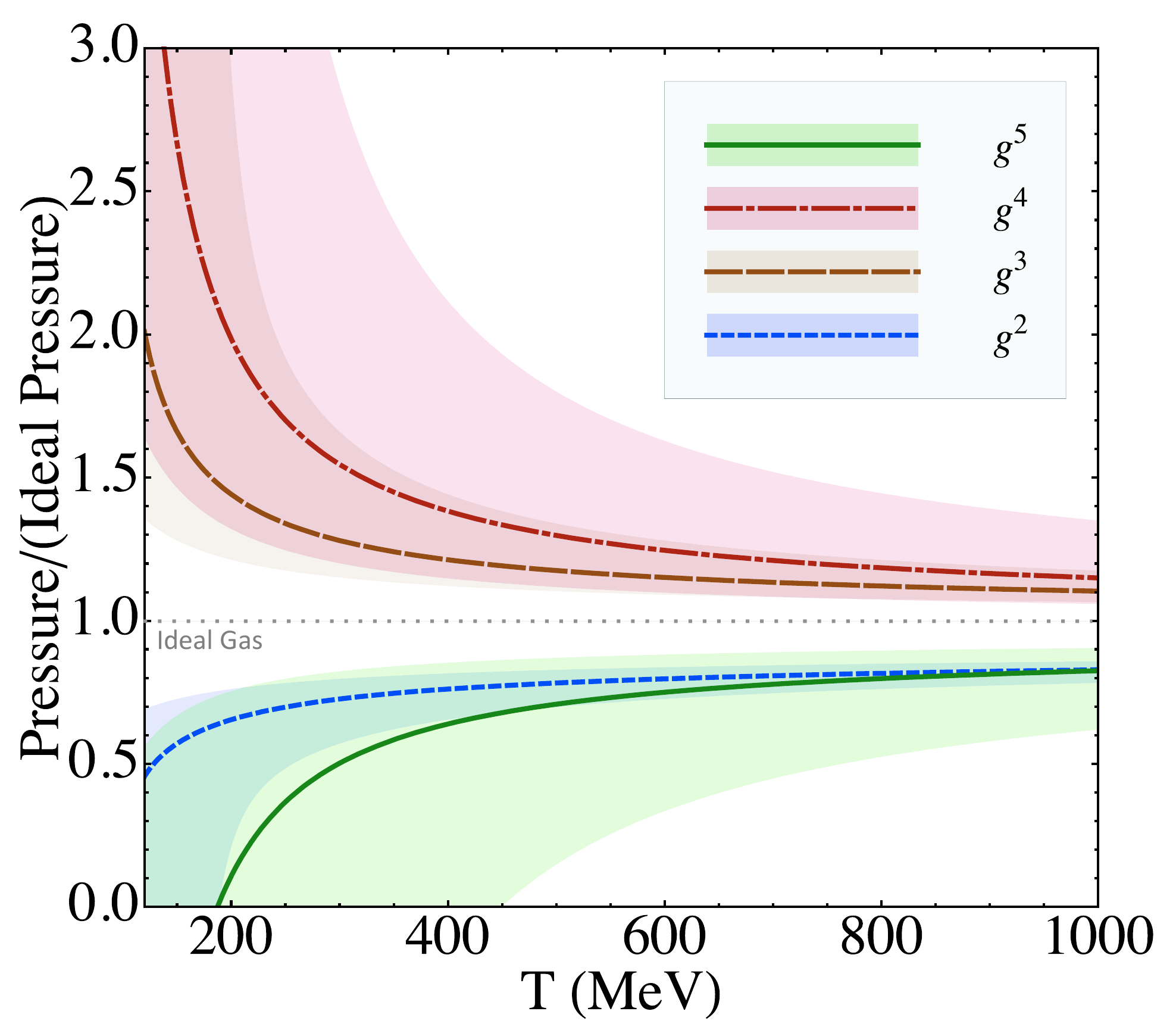}
  \caption{Weak-coupling expansion of the normalized pressure in 
three-flavor massless QCD
as a function of the temperature $T$. See main text for details.}
\label{weak}
\end{figure}

In Fig.~\ref{weak}, we show the weak-coupling expansion of the
QCD pressure with $N_f=3$ normalized to that of an ideal gas of 
quarks and gluons
through order $g^2$, $g^3$, $g^4$, and $g^5$.
The curves are obtained by using the strong coupling constant
$g(\Lambda)$ evaluated at the renormalization scale $\Lambda=2\pi T$.  
The bands are obtained by varying the renormalization scale by a factor of two 
around this central value.
As successive terms in the weak-coupling expansion are added,
the predictions for the pressure fluctuate wildly and the 
sensitivity to the renormalization scale $\Lambda$ grows.
Due to asymptotic freedom, the weak-coupling expansion does converge
for sufficiently high temperatures.
However, this is only the case for temperatures many orders of magnitude
larger than the critical temperature $T_c$ for deconfinement.
For example the order-$g^3$ term is smaller than 
the order-$g^2$ only if $T$ is larger than approximately $10^5T_c$.
We note in passing that the poor convergence is not specific to QCD,
it is a generic feature of hot quantum field theories. 
This together with the large $g^3$ corrections suggests that the
problem is in the soft sector of the theory, i.e. momenta on the order
of $gT$, and is not related to the breakdown of perturbation theory
due to infrared divergences at four loops (Linde's problem)~\cite{linde}.
There are several ways of reorganizing the perturbative 
series~\cite{review1,review2}, here I will focus on 
hard-thermal-loop perturbation theory (HTLpt) and dimensional 
reduction (DR).
The results presented in this talk are published
in the papers~\cite{leg1,syl,3l1,3l2}.
\\

\section{Reorganization of the perturbative series}
\subsection{Screened perturbation theory}
\begin{figure}[htb]
  \includegraphics[height=.05\textheight]{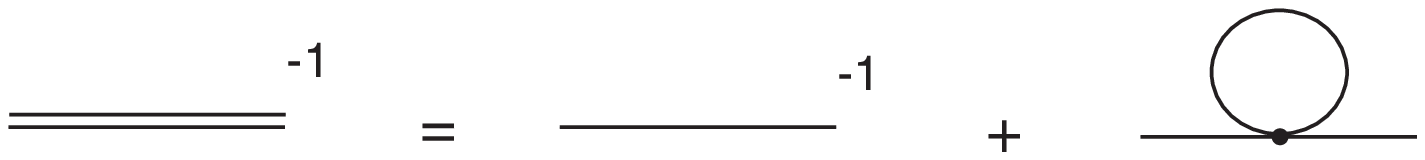}
  \caption{Inverse propagator to one-loop order.}
\label{1loop}
\end{figure}

Before we discuss hard-thermal-loop perturbation theory in some detail, 
it is useful to take a step back and discuss resummation in 
a self-interacting scalar field theory. The starting point is the 
Euclidean Lagrangian for massless $\phi^4$-theory
\bqa
{\cal L}&=&
{1\over2}(\partial_{\mu}\phi)^2+{g^2\over24}\phi^4\;.
\eqa
We are interested in calculating perturbative corrections to various
quantities, for example the two-point function.
Fig.~\ref{1loop} shows diagrammatically the inverse propagator to one-loop
order. 
\\
Calculating the one-loop diagram, the inverse propagator can be
written as
\bqa
\Delta^{-1}&=&P^2+{g^2\over24}T^2\;,
\eqa
where $P=(P_0=\omega_n,{\bf p})$.
For hard momenta, i.e. for $\omega_n\neq0$ or ${\bf p}\sim T$, we see
that the correction $\sim g^2T^2$ is a perturbative correction at weak coupling.
However, for soft momenta, i.e. for $\omega_n=0$ or ${\bf p}\sim gT$,
the correction is as large as the tree-level term.
This is the simplest example of a hard thermal loop (HTL)~\cite{eric}, namely
a loop correction which is dominated by hard loop momenta on the order
$T$ and is as large as a tree-level term. In this case, the HTL
is a simple local mass term. The fact that this term is as large
as the tree-level term suggests that we need to reorganize perturbation theory
at high temperature.

\begin{figure}[htb]
  \includegraphics[height=.06\textheight]{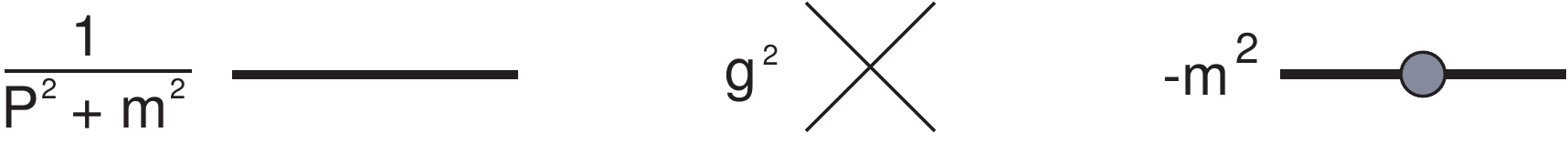}
  \caption{Feynman rules in screened perturbation theory.}
\label{rules}
\end{figure}

Screened perturbation theory (SPT)~\cite{spt0,chiku,spt3,spt4,spt5}
is one way of reorganizing
the perturbative series, which was inspired by variational perturbation
theory~\cite{vpt1,vpt2,vpt3}. It is 
defined by writing the Lagrangian as follows
\bqa
{\cal L}_0&=&
{1\over2}(\partial_{\mu}\phi)^2+{1\over2}m^2\phi^2,
\\
{\cal L}_{\rm int}&=&
-{1\over2}m^2\phi^2+{g^2\over24}\phi^4\;.
\eqa
The perturbative expansion is then an expansion around an ideal gas of
massive particles. In other words, we incorporate a (thermal)
mass to all loop orders via the propagator.
In order to avoid overcounting of Feynman diagrams, we
need to subtract the quadratic term and consider it as an interaction
on the same footing as the quartic interaction. The Feynman rules 
for SPT are shown in Fig.~\ref{rules}
\\
\\
Using a massive propagator, all infrared divergences are screened and 
one can, in principle, calculate Feynman diagrams at any order in 
screened perturbation theory. However, at this stage the mass parameter
$m$ in the Lagrangian is arbitrary and in order to complete
a calculation in SPT, we need a prescription for it.
I will return to this point later in my talk.

\subsection{Hard-thermal-loop perturbation theory}
Hard-thermal-loop perturbation theory is a generalization of
SPT to gauge theories and was developed by 
Andersen, Braaten, and Strickland over a decade ago~\cite{andersen1}.
In gauge theories one cannot simply add and subtract a local mass
term for the gluons as this would violate gauge invariance.

\begin{figure}[htb]
  \includegraphics[height=.06\textheight]{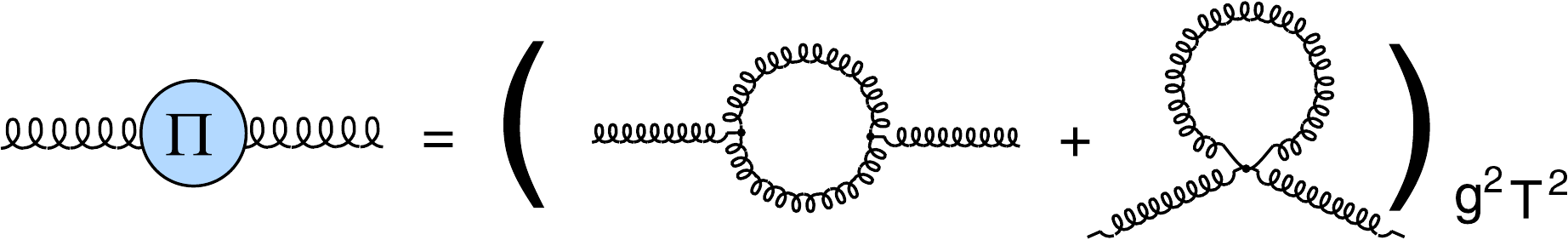}
  \caption{Hard thermal loop in nonabelian gauge theory.}
\label{htl1}
\end{figure}

\begin{figure}[htb]
  \includegraphics[height=.10\textheight]{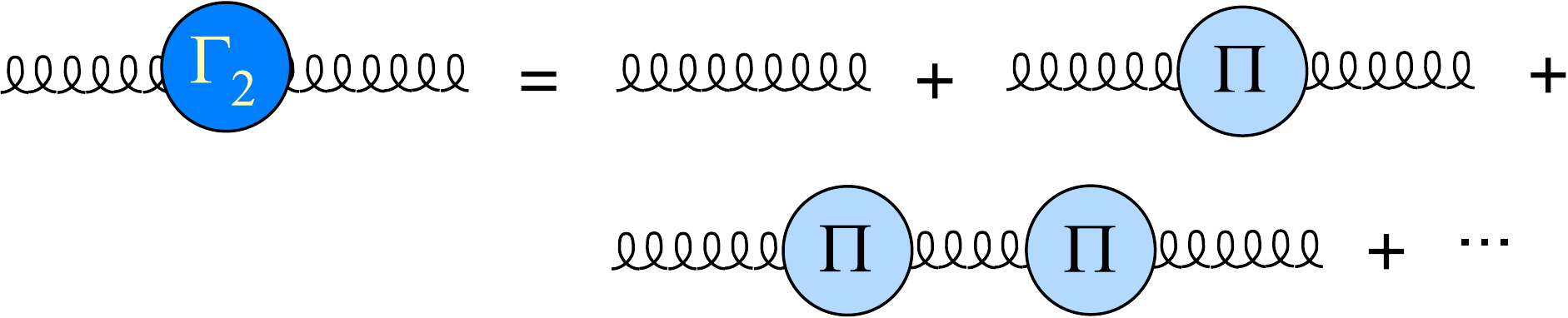}
  \caption{Inverse propagator to one-loop order.}
\label{1loopsum}
\end{figure}

Looking more carefully at the gluon self-energy function at one-loop order, 
one realizes that there are contributions (see Fig.~\ref{htl1})
which are as large as the tree-level term for soft external momenta. 
The corresponding loop integral is again dominated by hard momenta
and is another example of a hard thermal loop.
The contribution to the self-energy function $\Pi$
from the one-loop diagram is then used
to construct an effective two-point function, as shown in Fig.~\ref{1loopsum}.
\\

\begin{figure}[htb]
  \includegraphics[height=.14\textheight]{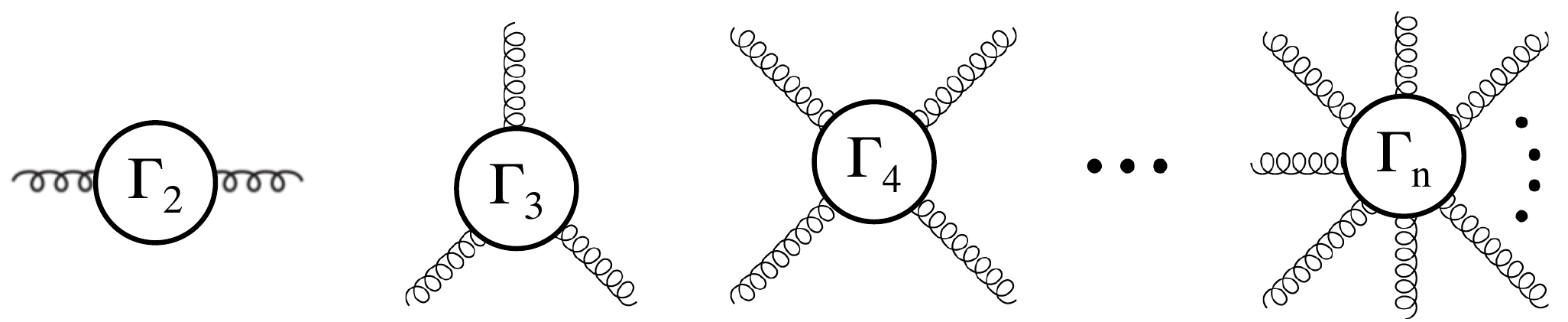}
  \caption{Effective $n$-point functions in gauge theories.}
\label{vertices}
\end{figure}

It turns out that not only the two-point function receives loop
corrections that are as a large as the tree-level contribution, but
also higher $n$-point functions do. Thus, we need to use 
effective vertices together with effective propagators, see Fig.~\ref{vertices}.
This is essential in order to maintain gauge invariance. 
\\
There is a nonlocal effective action that generated all the 
hard-thermal-loop $n$-point functions~\cite{wong,efflag1}.
It can be written in a manifestly gauge invariant form and reads
\bqa
 {\cal L}_{\rm HTL}&=&
-\frac{1}{2}(1-\delta)
 m_D^2 {\rm Tr}\lb G_{\mu\alpha}\left\langle\frac{y^\alpha y_\beta}{(y\cdot\! D)^2}
\right\rangle_{\!\hat{\bf y}} G^{\mu\beta}\rb 
+
(1-\delta)i m_q^2\bar\psi\gamma^\mu
\left\langle\frac{y_\mu}{y\cdot\! D}\right\rangle_{\!\hat{\bf y}}\psi\;,
\label{htl_lag}
\eqa
where $G_{\mu\nu}$ is the field strength, $D$ is the covariant derivative,
and $m_D$ and $m_q$ are the Debye screening mass and fermion mass
parameters.
Moreover, $y=(1,{\hat{\bf y}})$ is lightlike four-vector with ${\hat{\bf y}}$
being a unit three-vector.
The
bracket $\langle\rangle_{\!\hat{\bf y}}$ indicates an angular average over
the directions of ${\hat{\bf y}}$.
The QCD Lagrangian is then reorganized by writing
\bqa
{\cal L}&=&\left({\cal L}_{\rm QCD}+{\cal L}_{\rm HTL}\right)
\big|_{g\rightarrow g\sqrt{\delta}}+\Delta{\cal L}_{\rm HTL}\;,
\label{totally}
\eqa
where the QCD Lagrangian in Minkowski space is
\bqa
{\cal L}_{\rm QCD}
&=&
-\frac{1}{2}{\rm Tr}[G_{\mu\nu}G^{\mu\nu}]+i\bar\psi\gamma^\mu D_{\mu}\psi
+{\cal L}_{\rm gh}+{\cal L}_{\rm gf}
+\Delta{\cal L}_{\rm QCD}\;.
\label{qcd_lag}
\eqa
The term $\Delta{\cal L}_{\rm QCD}$ contains counterterms necessary
to cancel the ultraviolet divergences in perturbative calculations, and
$\Delta{\cal L}_{\rm HTL}$ contains the counterterms necessary
to cancel additional ultraviolet divergences generated by HTLpt.
${\cal L}_{\rm gh}$ is the ghost term that depends on the
gauge-fixing term ${\cal L}_{\rm gf}$, however we emphasize that HTLpt
is a gauge invariant
framework by construction. It can be used to calculate
both static and dynamical quantities. We point out, however, that HTLpt
suffers from the same infrared problems related to the magnetic mass
as does perturbative QCD, cf. Linde's problem~\cite{nan}.
HTLpt systematically shifts the perturbative expansion from being around an 
ideal gas of massless particles to being around a gas of massive quasiparticles 
which are the appropriate physical degrees of freedom at high temperature 
and/or chemical potential.

Note that the HTLpt Lagrangian~(\ref{totally}) reduces to the 
QCD Lagrangian (\ref{qcd_lag}) if we set $\delta=1$.
The parameter $\delta$ is an expansion parameter and HTLpt is
defined by an expansion in powers of $\delta$ around $\delta=0$.
For example at leading order in $\delta$, $\delta^0$, HTLpt describes
free massive quasiparticles that include 
screening effects and Landau damping.
At higher orders in $\delta$, we include interactions among these 
quasiparticles.

\begin{figure}[htb]
\includegraphics[height=.40\textheight]{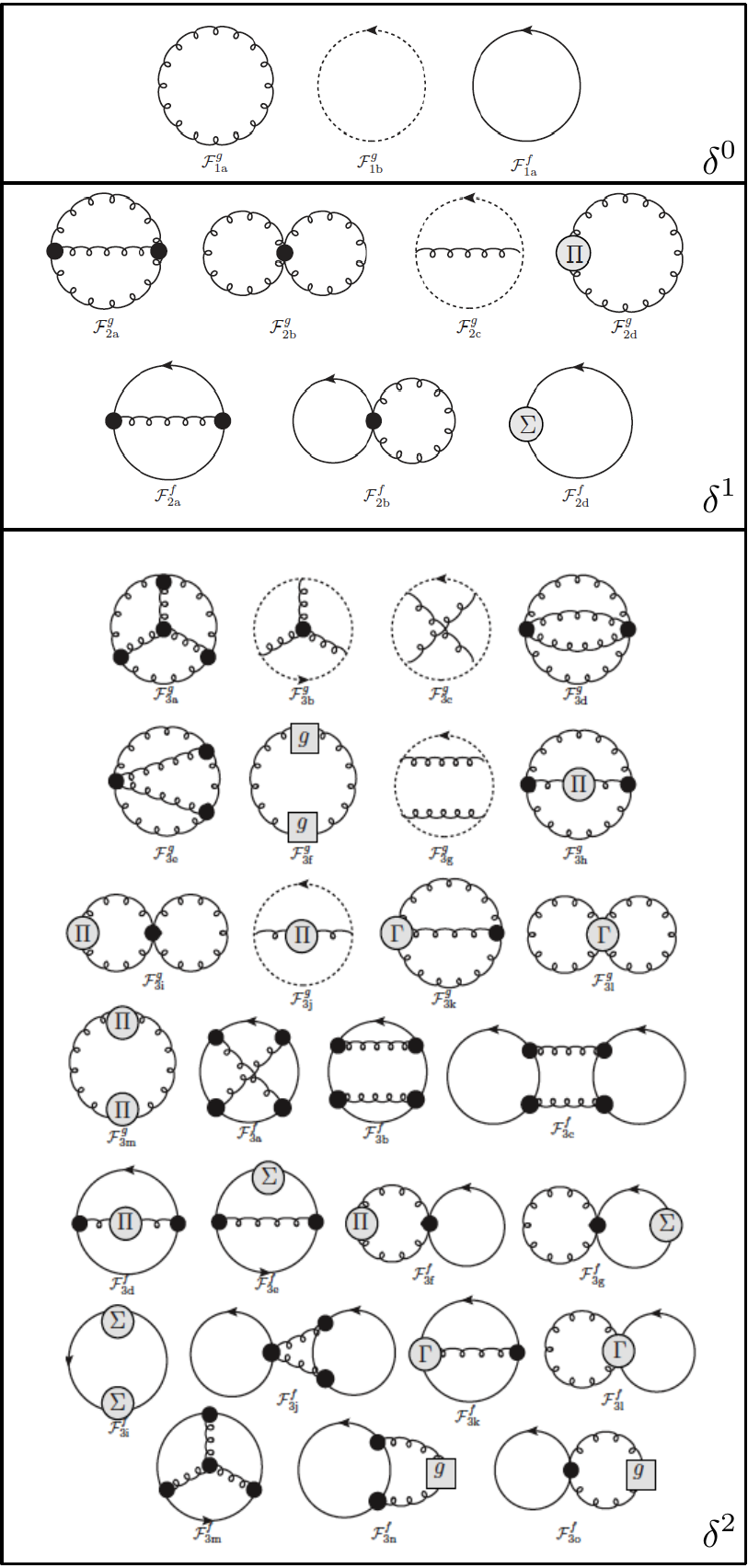}
\includegraphics[height=.30\textheight]{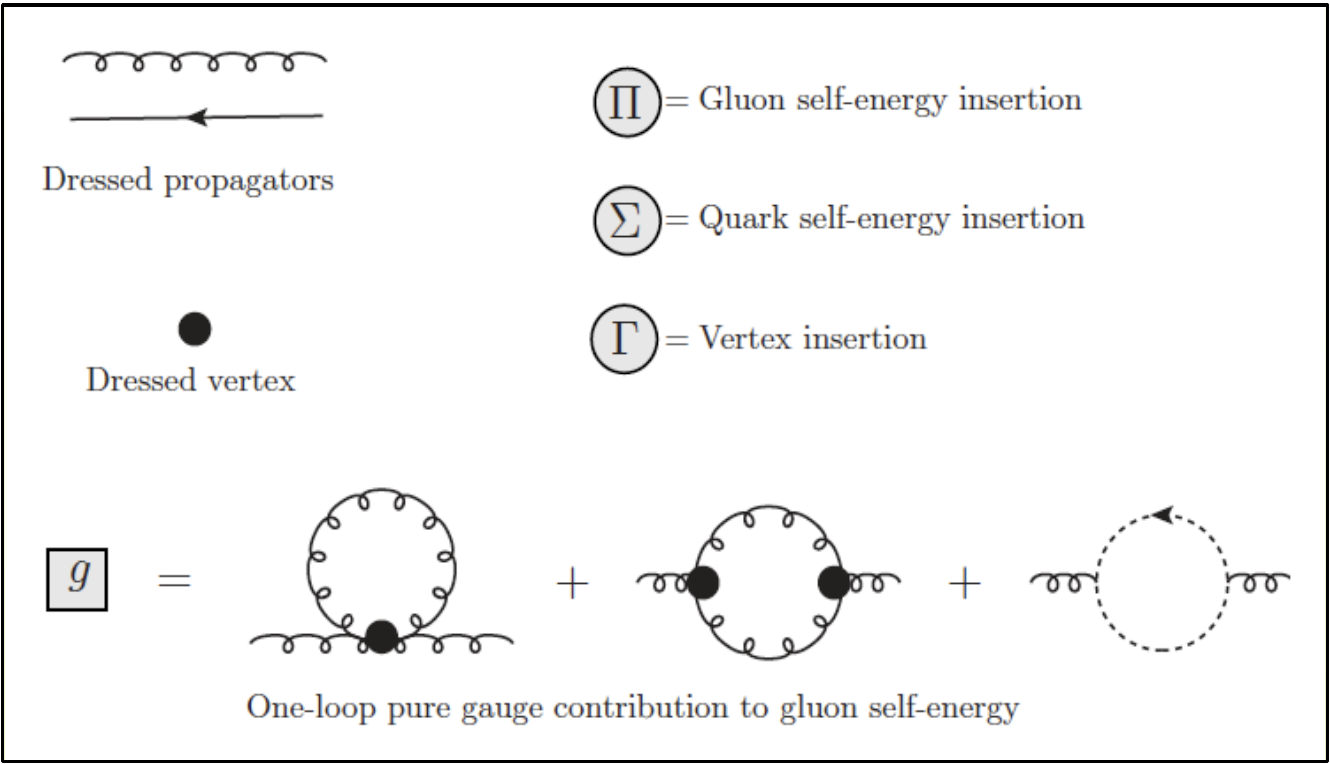}
\caption{Feynman diagrams that contribute to the free energy
through next-to-next-to-leading order (left) and various insertions
(right).}
\label{vacuumd}
\end{figure}

In Fig.~\ref{vacuumd} we show the diagrams contributing
to the pressure  at order
$\delta^0$, $\delta$, and $\delta^2$ (left). 
The different insertions are
shown to the right.  
\\
A three-loop calculation at finite temperature  
and with a separate quark chemical potential for each quark
yields~\cite{3l1,3l2}.
\begin{eqnarray}
\frac{\Omega_{\rm NNLO}}{\Omega_0}
&=& \frac{7}{4}\frac{d_F}{d_A}\frac{1}{N_f}
\sum\limits_f\lb1+\frac{120}{7}\hmu_f^2+\frac{240}{7}\hmu_f^4\rb
    -\frac{s_F\alpha_s}{\pi}\frac{1}{N_f}
\sum\limits_f\bigg[\frac{5}{8}\left(1+12\hat\mu_f^2\right)\left(5+12\hat\mu_f^2
\right)
    \nn
    &&-\frac{15}{2}\left(1+12\hat\mu_f^2\right)\hat m_D
-\frac{15}{2}\bigg(2\ln{\frac{\hat\Lambda}{2}-1
   -\aleph(z_f)}\Big)\hat m_D^3
      +90\hat m_q^2 \hat m_D\bigg]
\nn
&&+ \frac{s_{2F}}{N_f}\left(\frac{\alpha_s}{\pi}\right)^2
\sum\limits_f\bigg[\frac{15}{64}\bigg\{35-32\lb1-12\hmu_f^2\rb\frac{\zeta'(-1)}
      {\zeta(-1)}+472 \hat\mu_f^2+1328  \hat\mu_f^4\nn
      &&+ 64\Big(-36i\hat\mu_f\aleph(2,z_f)+6(1+8\hat\mu_f^2)
\aleph(1,z_f)+3i\hat\mu_f(1+4\hat\mu_f^2)\aleph(0,z_f)\Big)\bigg\}\nn
      &&- \frac{45}{2}\hat m_D\left(1+12\hat\mu_f^2\right)\bigg] \nn
&&+ \left(\frac{s_F\alpha_s}{\pi}\right)^2
      \frac{1}{N_f}\sum\limits_{f}\frac{5}{16}\Bigg[96\left(1+12\hat\mu_f^2
\right)\frac{\hat m_q^2}{\hat m_D}
     +\frac{4}{3}\lb1+12\hmu_f^2\rb\lb5+12\hat\mu_f^2\rb
      \ln\frac{\hat{\Lambda}}{2}\nn
    && +\frac{1}{3}+4\gamma_E+8(7+12\gamma_E)\hat\mu_f^2+112\mu_f^4
-\frac{64}{15}\frac{\zeta^{\prime}(-3)}{\zeta(-3)}-
   \frac{32}{3}(1+12\hat\mu_f^2)\frac{\zeta^{\prime}(-1)}{\zeta(-1)}\nn
   &&-    96\Big\{8\aleph(3,z_f)+12i\hat\mu_f\aleph(2,z_f)-2(1+2\hat\mu_f^2)
\aleph(1,z_f)-i\hat\mu_f\aleph(0,z_f)\Big\}\Bigg] \nn \nonumber
%
&&+ \left(\frac{s_F\alpha_s}{\pi}\right)^2
      \frac{1}{N_f^2}\sum\limits_{f,g}\left[\frac{5}{4\hat m_D}
\left(1+12\hat\mu_f^2\right)\left(1+12\hat\mu_g^2\right)
     +90\Bigg\{ 2\left(1 +\gamma_E\right)\hat\mu_f^2\hat\mu_g^2
      \right.\nn
        &&-\Big\{\aleph(3,z_f+z_g)+\aleph(3,z_f+z_g^*)+ 4i\hat\mu_f
\left[\aleph(2,z_f+z_g)+\aleph(2,z_f+z_g^*)\right]-4\hat\mu_g^2\aleph(1,z_f)\nn
       &&
       -(\hat\mu_f+\hat\mu_g)^2\aleph(1,z_f+z_g)- (\hat\mu_f-\hat\mu_g)^2
\aleph(1,z_f+z_g^*)-4i\hat\mu_f\hat\mu_g^2\aleph(0,z_f)\Big\}\Bigg\}\nn
       &&-\left.\frac{15}{2}\lb1+12\hat\mu_f^2\rb\lb2\L-1-\aleph(z_g)
\rb  \hat m_D\right]
\nn
&&+ \left(\frac{c_A\alpha_s}{3\pi}\right)\left(\frac{s_F\alpha_s}{\pi N_f}\right)
\sum\limits_f\Bigg[\frac{15}{2\hat m_D}\lb1+12\hmu_f^2\rb
     -\frac{235}{16}\Bigg\{\bigg(1+\frac{792}{47}\hat\mu_f^2+\frac{1584}{47}\hat
\mu_f^4\bigg)\ln\frac{\hat\Lambda}{2}
     \nonumber\\
    &&-\frac{144}{47}\lb1+12\hmu_f^2\rb\ln\hat m_D+\frac{319}{940}
\left(1+\frac{2040}{319}\hat\mu_f^2+\frac{38640}{319}\hat\mu_f^4\right)
   -\frac{24 \gamma_E }{47}\lb1+12\hat\mu_f^2\rb
\nonumber\\
    &&
   -\frac{44}{47}\lb1+\frac{156}{11}\hmu_f^2\rb\frac{\zeta'(-1)}{\zeta(-1)}
    -\frac{268}{235}\frac{\zeta'(-3)}{\zeta(-3)}
   -\frac{72}{47}\Big[4i\hat\mu_f\aleph(0,z_f)+\left(5-92\hat\mu_f^2\right)
\aleph(1,z_f)
    \nonumber\\
    &&+144i\hmu_f\aleph(2,z_f)
   +52\aleph(3,z_f)\Big]\Bigg\}+90\frac{\hat m_q^2}{\hat m_D}+\frac{315}{4}\Bigg
\{\lb1+\frac{132}{7}\hmu_f^2\rb\L
\nonumber\\
   &&+\frac{11}{7}\lb1+12\hmu_f^2\rb\gamma_E+\frac{9}{14}\lb1+\frac{132}{9}
\hmu_f^2\rb
+\frac{2}{7}\aleph(z_f)\Bigg\}\hat m_D 
\Bigg]
+ \frac{\Omega_{\rm NNLO}^{\rm YM}}{\Omega_0} \, ,
\label{finalomega}
\end{eqnarray}
where the sums over $f$ and $g$ include all quark flavors, 
$z_f = 1/2 - i \hat{\mu}_f$, $\Omega_{\rm NNLO}^{\rm YM}$ is the pure-glue 
contribution~\cite{3loopglue2}, 
$\Omega_0=-\mbox{$d_A\pi^2T^4/45$}$,
$\hat{m}_D=m_D/(2\pi T)$ etc, and $\alpha_s=g^2/(4\pi)$.
Moreover $\aleph(z)=\Psi(z)+\Psi(z^*)$, where $\Psi(z)$
is the digamma function and 
$\aleph(n,z)=\zeta^{\prime}(-n,z)+(-1)^{n+1}\zeta^{\prime}(-n,z^*)$
where $\zeta^{\prime}(x,y)=\partial_x\zeta(x,y)$.
$c_A=N_c$, $d_A=N_c^2-1$, $s_F=\mbox{$1\over2$}N_f$, $s_{2F}=C_Fs_F$ with
$C_F=\mbox{$N_c^2-1\over2N_c$}$.
Note that there are two renormalization scales $\Lambda_g$
and $\Lambda_q$, we use former in the purely gluonic graphs and the latter
in all other graphs. In this way the susceptibilities vanish in
the limit $N_f\rightarrow0$. In addition, the thermodynamic 
potential produces the correct ${\cal O}(g^5)$
weak-coupling result if expanded in a strict power series in $g$.

\subsection{Dimensional reduction}
\begin{figure}[htb]
\includegraphics[height=.15\textheight]{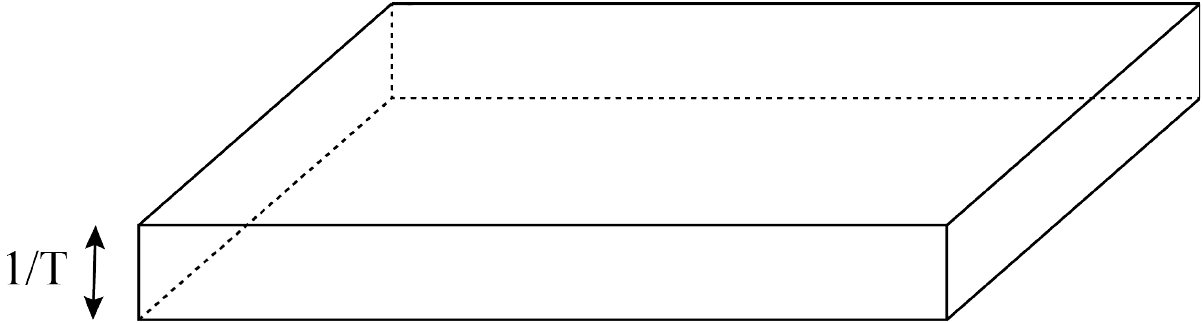}
\caption{Dimensional reduction in hot field theories.}
\label{dimred12}
\end{figure}

In scalar field theory, we have seen that the resummed effective propagator
is of the form $1/(P^2+m^2)$, where $m$ is on the order $gT$.
There are two momentum scales in scalar theory (and in QED), namely the
hard scale $T$ and the soft scale $gT$ ($eT$, where $e$ is the electric charge).
At weak coupling, the term $m^2$ is a perturbative correction
to the propagator for the nonzero Matsubara modes, while it is essential
to include for the static mode $P_0=0$. At weak coupling, one therefore
has to well separated mass scales and one can integrate out the
nonzero Matsubara frequencies to obtain an effectice 
three-dimensional field theory for the zeroth Matsubara 
mode~\cite{eric2,laine}. The idea is shown in Fig.~\ref{dimred12}, where
the imaginary time dimension is $1/T$ and so the system becomes effectively
three-dimensional at high temperature. 
\\

The scale $T$ can be integrated out perturbatively and in the case of
scalar $\phi^4$-theory, the effective three-dimensional 
Lagrangian becomes
\bqa
{\cal L}_{\rm eff}&=&{1\over2}(\nabla\phi)^2+{1\over2}m_3^2\phi^2
+{\lambda\over24}\phi^4+\delta{\cal L}\;,
\eqa
where $\delta{\cal L}$ contains higher-order operators.
The parameters of the effective Lagrangian are functions of $T$,
$g^2$, and a renormalization scale $\Lambda$.
For example $m_3^2={g^2\over24}T^2$ and $\lambda=g^2T$ to leading order
in $g^2$. In QED, dimensional reduction can be carried out the same
way and one obtains an effective three-dimensional theory for a
massive scalar field $A_0$, which can be identified with the zeroth
component of the original gauge field, and a gauge field $A_{i}$, which can 
identified with the spatial components of the original gauge 
field~\footnote{The fermions are massive since $\omega_n=(2n+1)\pi T$
so they are integrated out as well.}.
In QCD, it is somewhat more complicated as we have three momentum scales,
namely the hard scale $T$, soft scale $gT$, and the supersoft scale
$g^2T$. At weak coupling, we then have three well separated momentum
scales and we can successive integrate out the scales $T$ and $gT$.
Integrating out the scale $T$, we obtain an effective three-dimensional
theory, Electrostatic QCD (EQCD), whose Lagrangian is
\bqa
{\cal L}_{\rm EQCD}&=&{1\over2}{\rm Tr}\left[G_{ij}\right]
+{\rm Tr}\left[(D_iA_0)^2\right]
+M_E^2{\rm Tr}\left[A_0^2\right]
+i\xi{\rm Tr}\left[A_0^3\right]
+\lambda_E{\rm Tr}\left[A_0^4\right]+\delta{\cal L}
\;,
\eqa
where $G_{ij}$ is the nonabelian field strength tensor and
$A_0$ is a scalar field in the adjoint representation of the gauge group
$SU(3)$. Moreover, $m_E$ is the color electric screening mass,
$\lambda_E$ is the scalar self-coupling,  and 
$\xi\propto\sum_f\mu_f$~\cite{dimred}.
The term $\delta{\cal L}$ contains all higher-order operators that start
to contribute to the pressure beyond order $g^6$.
We note in passing that EQCD breaks the $SU(N_c)$ center symmetry
of four-dimensinonal QCD, which 
can be remedied by formulating the theory in terms
of Wilson loop type variables instead of the $A_0$ field~\cite{poly1,poly2}.
However, for high temperatures, this is of little consequence
as the probability of thermal fluctuations 
crossing the barriers that separate the trivial minimum
we are expanding about, and other minima is exponentially small.

Dimensional reduction has been carried out both at zero 
density~\cite{finns1,finns2} as well as at finite density~\cite{aleksi}
to order $g^6\ln (g)$.
The color electric screening mass is on the order $gT$, and as a result, we
integrate it out perturbatively to obtain a second effective field
theory (Magnetostatic QCD). 
This theory is plagued with infrared divergences in perturbation theory
and must be treated nonperturbatively using lattice simulations.
We can now write the pressure in QCD as a sum of 
contributions from the different momentum scales
\bqa
{\cal P}&=&{\cal P}_{\rm hard}+{\cal P}_{\rm soft}+{\cal P}_{\rm supersoft}\;.
\eqa
The contribution from the hard scale comes from dimensional reduction
(which can be identified with the unit operator in ${\cal L}_{\rm EQCD}$).
The contribution from the soft scale comes from calculations in 
using EQCD, while the contribution from the supersoft scale
comes from calculations in MQCD. However, MQCD contributes first at order
$g^6$, and as a result, 
we can ignore these contributions if we restrict ourselves
to order $g^5$.

\section{Numerical results}
We next present our numerical results. Physical quantities
such as the pressure and susceptibilities depend on the mass parameters
$m_D$ and $m_f$ as well as the running coupling $\alpha_s$ and
the renormalization scales $\Lambda_g$ and $\Lambda_q$
We fixed the scale $\overline{\rm MS}$ by requiring that
$\alpha_s(1.5{\rm GeV})=0.326$, which is obtained from lattice 
measurements~\cite{runbaza}. Using one-loop running this gives
$\overline{\rm MS}=176$ MeV, while two-loop running gives
$\overline{\rm MS}=283$ MeV.
We used one-loop running for HTLpt and two-loop running for DR.
We take the central values $\Lambda_g=2\pi T$ and 
$\Lambda_q=2\pi\sqrt{T^2+\mu^2/\pi^2}$ in our HTL calculations.
In our DR calculations, we used a central value of 
$\Lambda=1.445\times 2\pi T$ for $\mu_q=0$,
which is based on the principle of fastest apparent convergence.
This can be generalized to nonzero density, see Ref.~\cite{syl}
for details.
In all plots, the thick lines indicate the results obtained using these
central values and the bands are obtained by 
varying the scales by a factor of two.
For the numerical results presented in this talk, 
we use $c_A=N_c$ and $N_f=3$.

As mentioned before, one needs to give a prescription for the 
mass parameters $m_D$ and $m_f$. One can think of a variational 
prescription such as 
\bqa
{\partial{\cal F}\over\partial m_D^2}&=&0\;,
\eqa
where ${\cal F}$ is the free energy. 
At one-loop order in HTLpt, the only solution is the trivial solution.
At higher orders, this equation has complex solution for some values
of the coupling.  We therefore decided to use value $m_f=0$ for
the fermion mass parameter 
and the Debye mass $m_D$ given by the two-loop expression
for the mass parameter in EQCD~\cite{aleksi},
\bqa
\tiny
\nonumber
\hat m_D^2&=&\frac{\alpha_s}{3\pi} \Biggl\{c_A
+\frac{c_A^2\alpha_s}{12\pi}\lb5+22\gamma_E+22\Lg\rb +
\frac{1}{N_f} \sum\limits_{f}
\Biggl[ s_F\lb1+12\hmu_f^2\rb
 +\frac{c_As_F\alpha_s}{12\pi}\bigg(
\lb9+132\hmu_f^2\rb
\\ && \nonumber
+22\lb1+12\hmu_f^2\rb\gamma_E
+2\lb7+132\hmu_f^2\rb\L+4\aleph(z_f)\bigg)
\\ &&
+\frac{s_F^2\alpha_s}{3\pi}\lb1+12\hmu_f^2\rb\lb1-2\L+\aleph(z_f)\rb
 -\frac{3}{2}\frac{s_{2F}\alpha_s}{\pi}\lb1+12\hmu_f^2\rb \Biggr] \Biggr\}\;.
\eqa
When we evaluate various physical quantities either in HTLpt or DR,
it is essential that we do not expand the mass parameter $m_D$
in powers of $g$, but keep the full $g$-dependence to resum
the perturbative series.

\subsection{Pressure}

\begin{figure}
\includegraphics[height=.30\textheight]{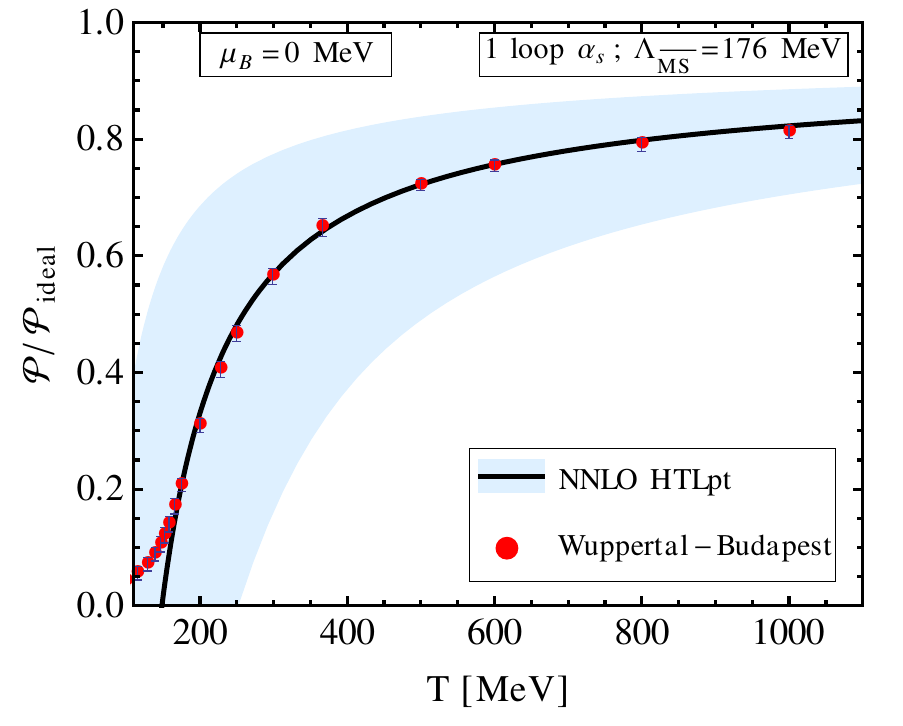}
\includegraphics[height=.30\textheight]{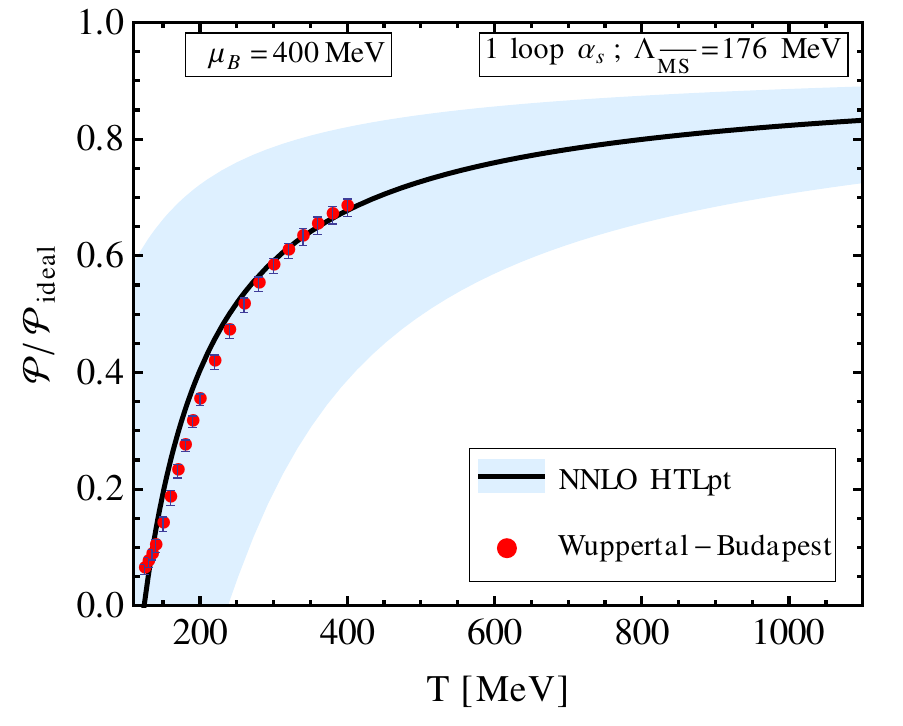}
\caption{The pressure normalized to the pressure of an ideal gas of massless
particles as a function of temperature $T$. Lattice data
are from ~\cite{borsap} and~\cite{borsapmu}, respectively.}
\label{presfig}
\end{figure}

In Fig.~\ref{presfig}, we show the normalized pressure as a function of 
$T$ for $\mu_B=0$ (left) and $\mu_B=400$ MeV (right).
The solid lines are obtained by using the
central value $\Lambda_g=2\pi T$ 
and $\Lambda_q=2\pi\sqrt{T^2+\mu^2/\pi^2}$ 
for the renormalization scales. The blue bands
are obtained by varying this scale around the central value by a factor
of two. The lattice data are from the Budapest-Wuppertal 
collaboration~\cite{borsap,borsapmu}. The agreement with the
central line and the lattice data is remarkable all the way down
to approximately $T=200$ MeV. Since HTLpt is a perturbative approach that
does not incorporate the $SU(N_c)$ center symmetry, there is no
reason to expect agreement with the data at temperatures close to $T_c$
and the agreement seen may be fortuitous.
\\

\subsection{Susceptibilities}
We next turn to the quark and baryon susceptibilities.
The pressure is a function of the temperature $T$ and the quark chemical 
potentials $\mu_q$. Furthermore, we can expand the 
pressure as a Taylor series in powers of ${\mu_q/T}$ around zero chemical
potential:
\bqa
{{\mathcal P}\over T^4}&=&{{\cal P}_0\over T^4}+
\sum_{ijk}{1\over i!j!k!}
\chi_{ijk}\left({\mu_u\over T}\right)^i
\left({\mu_d\over T}\right)^j\left({\mu_s\over T}\right)^k
\eqa
where the coefficients are given by
\bqa
\chi_{ijk...}&=&
{\partial^{i+j+k+...}{\cal P}(T,{\bf\mu})
\over\partial\mu^i_u\partial\mu_d^j\partial\mu_s^k}\bigg|_{\mu_q=0}\;.
\eqa
Below, we will use the shorthand notation for the susceptibilities by 
specifying derivatives by a string of quark flavors in superscript form, 
e.g. $\chi^{uu}_2=\chi_{200}$, $\chi^{ds}_2=\chi_{011}$, 
$\chi^{uudd}_2=\chi_{220}$ etc. We will also interested in the 
baryon-number susceptibilities, which 
are defined by
\bqa
\chi^B_n&=&
{\partial^n{\cal P}\over\partial\mu^n_B}\bigg|_{\mu_B=0}\;.
\eqa
Using the relation $\mu_B=\mu_u+\mu_d+\mu_s$ and the chain rule
we can find relations between the quark and baryon susceptibilities.
For example, one finds
\nonumber
\bqa
\chi_2^B&=&{1\over9}\left[
\chi_2^{uu}+\chi_2^{dd}+\chi_2^{ss}+2\chi_2^{ud}+2\chi_2^{us}+2\chi_2^{ds}
\right]\;.
\eqa

\begin{figure}[htb]
\includegraphics[height=.30\textheight]{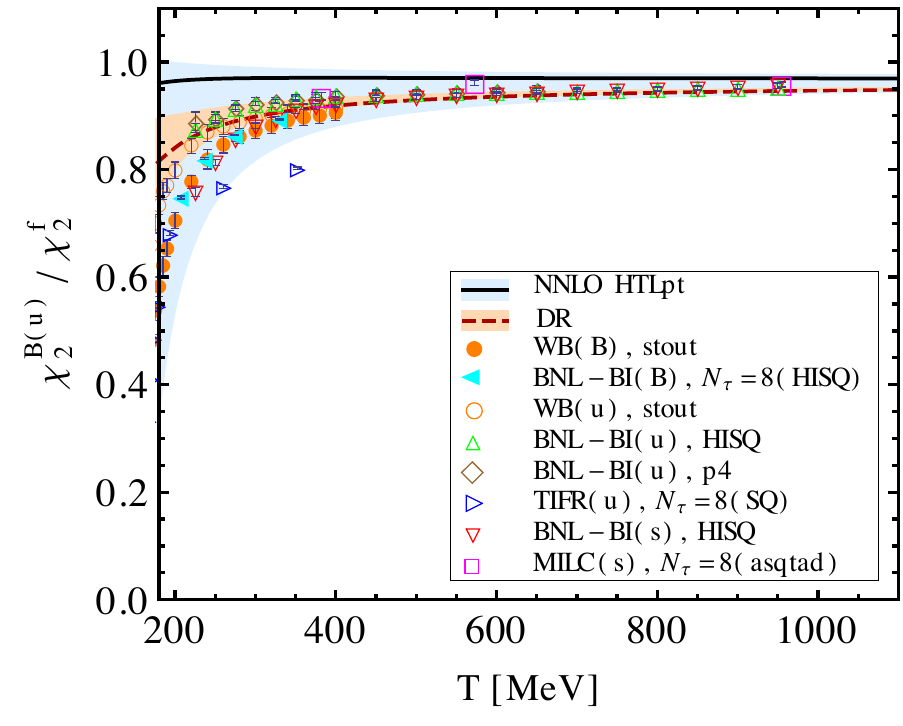}
\includegraphics[height=.30\textheight]{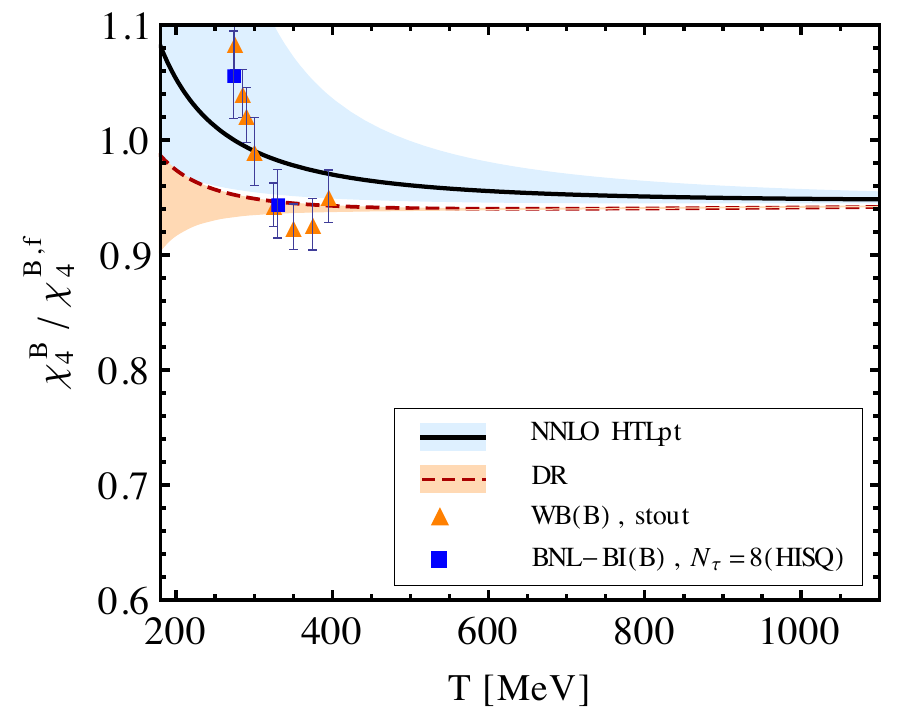}
\caption{The scaled second (left) and fourth-order (right)
baryon number susceptibility from HTLpt and DR
compared with various lattice data.
The lattice data labeled WB,
BNL-BI(B), BNL-BI(u,s), MILC, and TIFR come from 
Refs.~\cite{borsa1,baza1,baza2,milk,datta}.}
\label{suscfig}
\end{figure}

In Fig.~\ref{suscfig}, we compare the scaled second (left) and fourth-order 
(right) baryon number susceptibility 
from HTLpt and DR compared with various lattice data.
Both resummation schemes are in good agreement with the data.
We also notice that the band using DR is significantly smaller than
that obtained in HTLpt.

\begin{figure}[htb]
\includegraphics[width=0.45\textwidth]{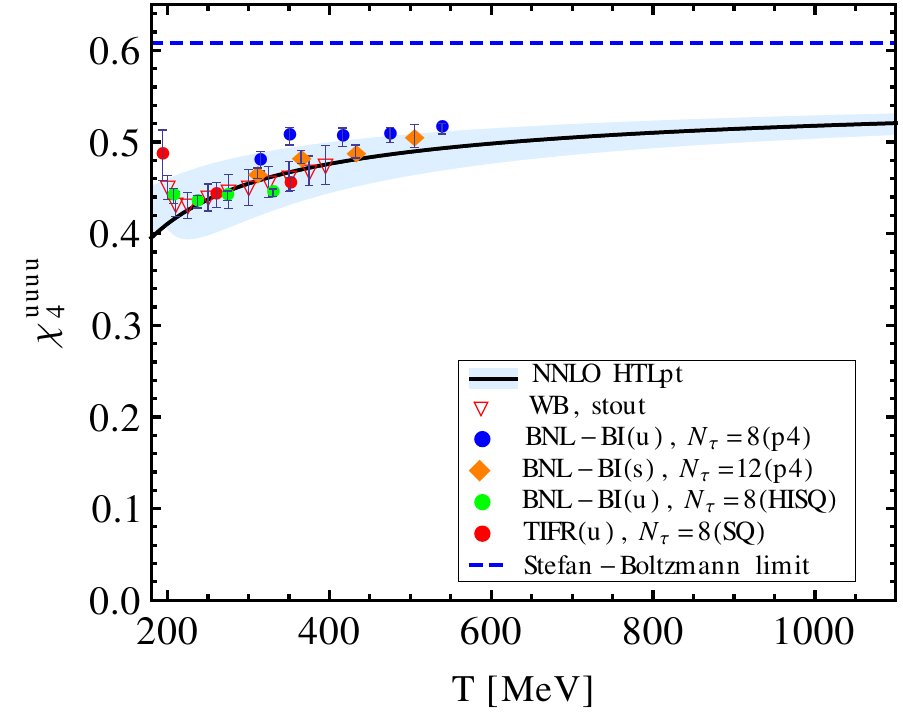}
\hspace{4mm}
\includegraphics[width=0.45\textwidth]{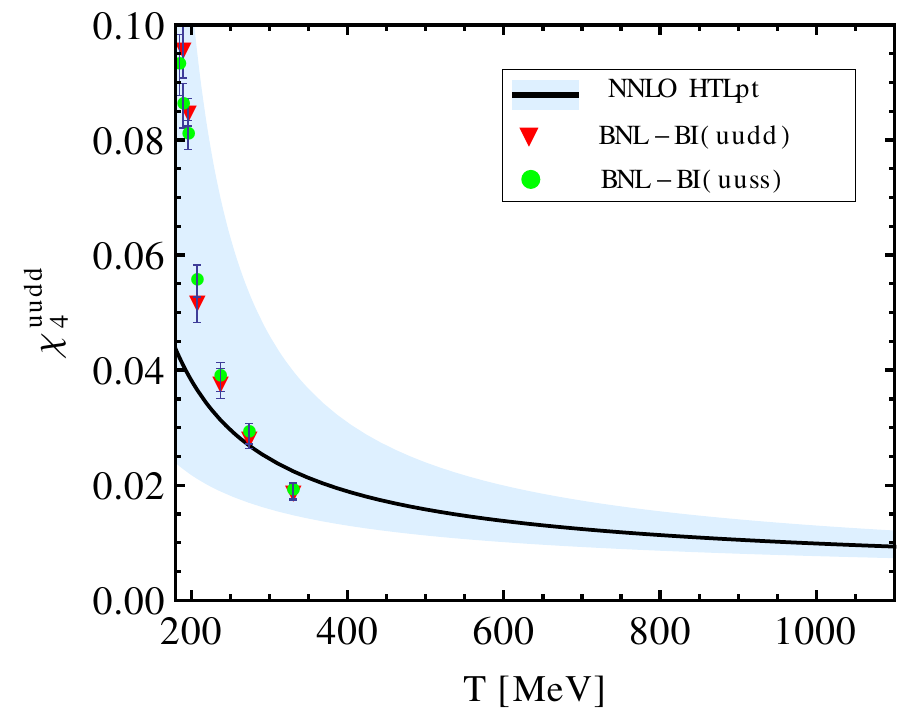}
\caption{
Comparison of the NNLO HTLpt fourth order diagonal single quark number 
susceptibility (left) and the only non-vanishing fourth order off-diagonal 
quark number susceptibility (right) with lattice data.  In the left figure the 
dashed blue line indicates the Stefan-Boltzmann limit for this quantity.  The 
data labeled BNL-BI(uudd), BNL-BI(u,s), BNL-BI(uuss), and TIFR come from 
Refs.~\cite{baza1,baza2,milk,datta}.}
\label{fig:qns}
\end{figure}

In Fig.~\ref{fig:qns}, we show
the NNLO HTLpt fourth order diagonal single quark number 
susceptibility (left) and the only non-vanishing fourth order off-diagonal 
quark number susceptibility (right) with lattice data.
Again the agreement between the HTLpt prediction and the lattice data
is good.

\subsection{Interaction measure and speed of sound}
\begin{figure}[htb]
  \includegraphics[height=.30\textheight]{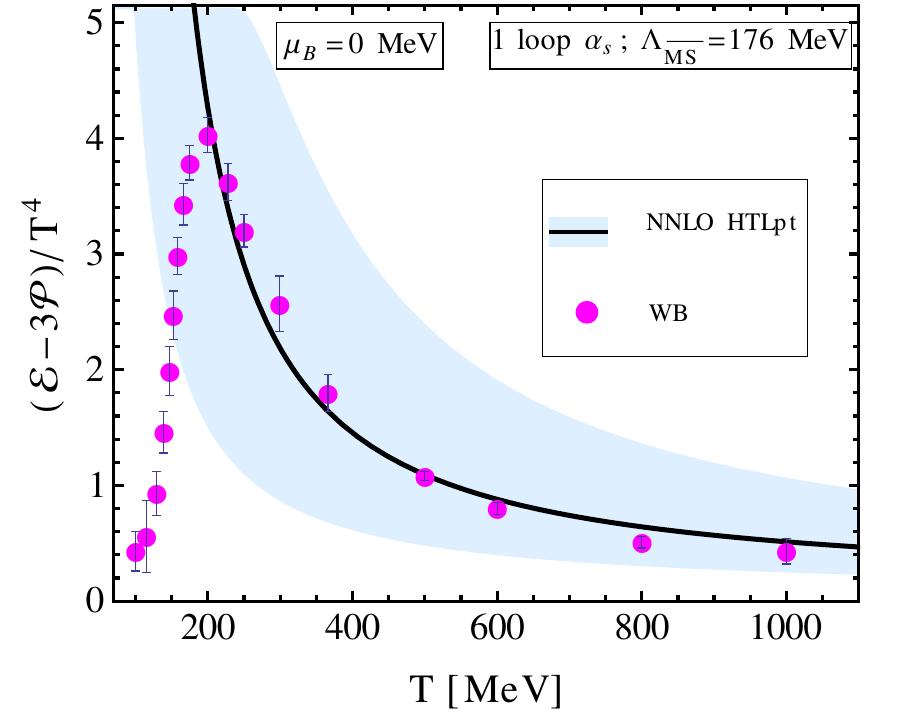}
  \includegraphics[height=.30\textheight]{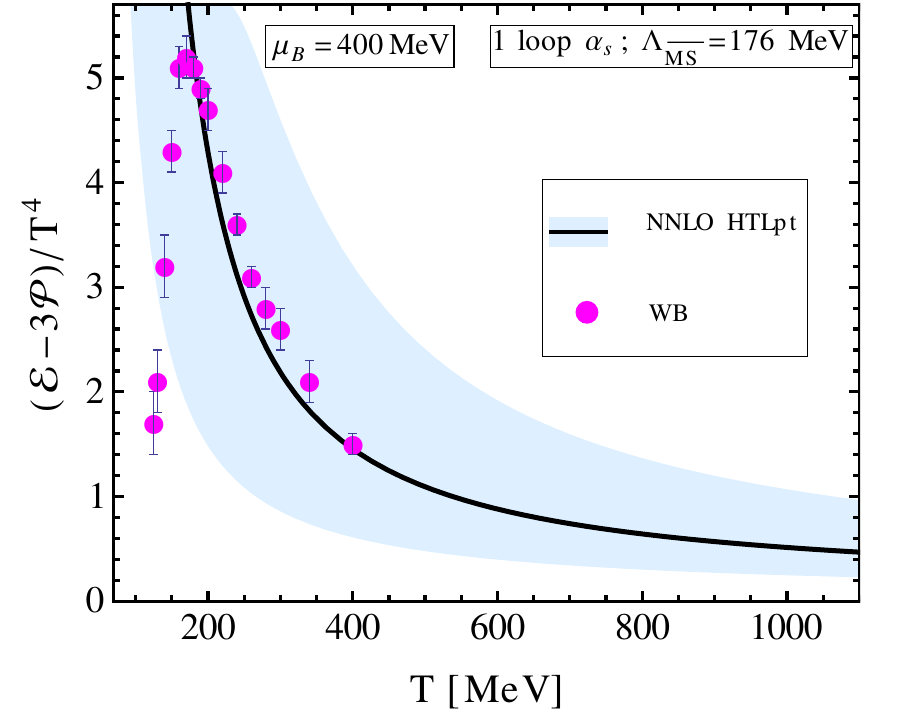}
  \caption{
Interaction measure ${\cal E}-3{\cal P}$ divided by $T^4$
as a function of temperature $T$
for $\mu_B=0$ (left) and $\mu_B=400$ MeV (right).
The $\mu_B=0$ lattice data are from \cite{borsap} and the $\mu_B=400$ MeV 
lattice data are from \cite{borsapmu}.} 
\label{traceanomaly}
\end{figure}

\begin{figure}[t]
  \includegraphics[height=.30\textheight]{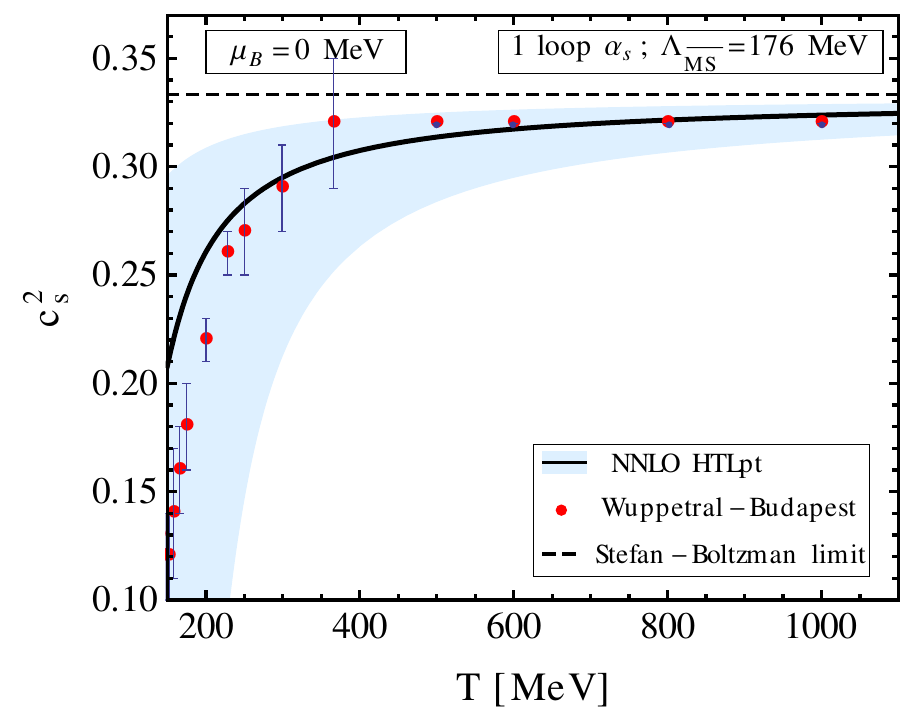}
  \includegraphics[height=.30\textheight]{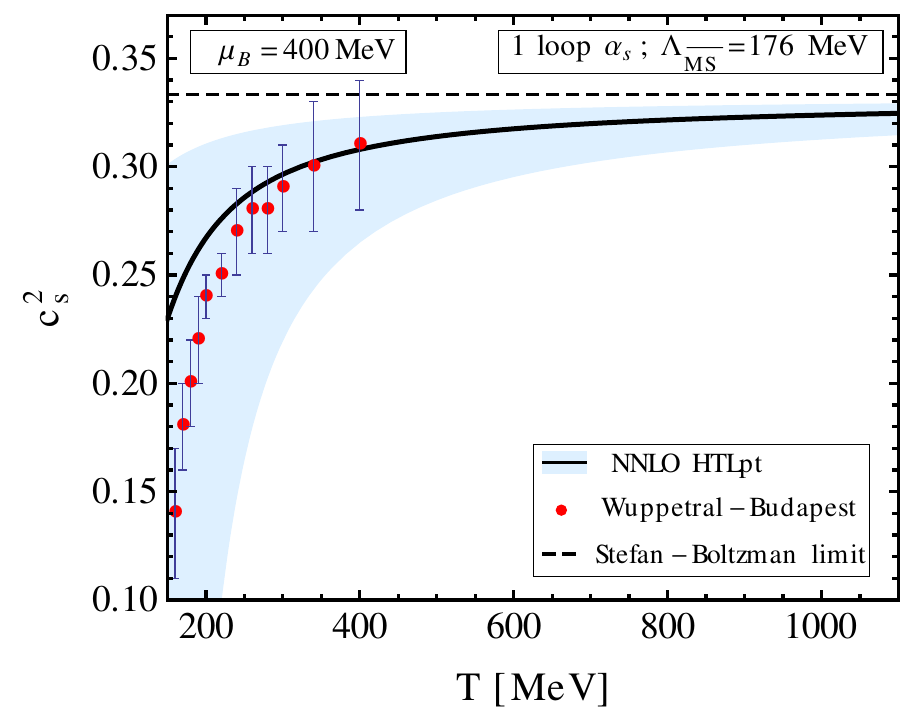}
\caption{Comparison of the $N_f=2+1$, $\mu_B=0$ (left) and $\mu_B=400$ MeV 
(right) NNLO 
HTLpt speed of sound squared with lattice data.
The $\mu_B=0$ lattice data are from \cite{borsap} and the $\mu_B=400$ MeV 
lattice data are from \cite{borsapmu}.
The dashed lines indicate the Stefan-Boltzmann limit.}
\label{cssq_1l}
\end{figure}

The interaction measure is given by
\bqa
{\cal E}-3{\cal P}\;,
\eqa
where ${\cal E}$ is the energy density.
Since  this quantity is the trace of the energy-momentum tensor
and vanishes for an ideal gas of massless particles,
it is also referred to as the trace anomaly.
In Fig.~\ref{traceanomaly}, we show the interaction measure
normalized by $T^4$ as a function of temperature $T$ for
$\mu_B=0$ (upper left panel) and 
$\mu_B=400$ MeV (upper right panel).
The lattice data are from the Wuppertal-Budapest 
collaboration~\cite{borsap,borsapmu}.
We see that the agreeement between the prediction from HTLpt
and lattice data is very good all the way down to $T\approx250$ MeV, which
is very near where the peak is located. For lower values of the temperature,
HTLpt fails completely, but this is not very surprising. 
There is no reason to expect that resummed perturbation theory 
works close to the QCD transition temperature.
\\
\\
Another quantity which is phenomenologically interesting is the speed of sound. 
The speed of sound squared is defined as
\bqa
c_s^2=\frac{\del{\cal P}}{\del{\cal E}} \, .
\eqa
In Fig.~\ref{cssq_1l} we plot the NNLO HTLpt speed of sound squared
for $\mu_B=0$ 
(left) and $\mu_B=400$ MeV (right) together with lattice data from 
Refs.~\cite{borsap} and \cite{borsapmu}.  
As we can see from this figure, there is quite good agreement between the NNLO 
HTLpt speed of sound and the lattice data when the central value of the scale 
is used.

\section{Summary and Outlook}
In this talk, I have presented results for QCD thermodynamic functions
using hard-thermal-loop perturbation theory and dimensional reduction.
We note that our results are completely analytic and
that, after we have chosen the renormalization scales
$\Lambda_g$ and $\Lambda_q$, there are no fit parameters. 
Comparing our results with available lattice data, they 
suggest that for temperatures above 
$250-500$ MeV (depending on the quantity we are looking at), 
the pressure, quark number susceptibilities, and other
quantities can be accurately described via resummed perturbation theory. 

Hard-thermal-loop perturbation theory represents a gauge-invariant
reorganization of the perturbative series. It
is formulated in Minkowski space and so can be 
used to calculate static and dynamical quantities alike.
The good agreement between HTLpt and lattice offers some hope that
applications to real-time quantities will be useful.

There are several directions for future work. One is to include finite
quark masses, another is to resum logarithms to reduce the scale
variation of our results. Extending our results to larger values of
$\mu_q$ is also of interest as there currently no other first principle
method is available.


\begin{theacknowledgments}
The speaker would like to thank the organizers for an interesting
and stimulating meeting. He also thanks his co-authors for
an enjoyable collaboration.
The authors thank S. Bors\'anyi, S. Datta, F. Karsch, S. Gupta, S. Mogliacci, 
P. Petreczky, and A. Vuorinen for useful discussions. 
\end{theacknowledgments}



\bibliographystyle{aipproc}   


\end{document}